\documentclass[preprint]{article}
\usepackage{authblk}
\usepackage{graphicx}
\usepackage{dcolumn}
\usepackage{bm}% bold math
\usepackage{amsmath}
\usepackage{amssymb}
\usepackage{multirow}
\usepackage{caption}
\usepackage{subcaption}
\usepackage{epstopdf}
\usepackage{hyperref}

\title{\bf Cosmological implications of LRS Bianchi type-I cosmological model in $f(T)$ gravity}
\author[1]{Shivangi Rathore\thanks{shivangirathore1912@gmail.com}}
\author[2]{S. Surendra Singh\thanks{ssuren.mu@gmail.com}}

\affil[1,2]{Department of Mathematics, National Institute of Technology Manipur, Imphal-795004,India.}

\begin{document}

\maketitle

\textbf{Abstract}:
 We perform the dynamical system analysis of the Locally Rotationally Symmetric (LRS) Bianchi type-I cosmological model in f(T) gravity in the presence of energy interaction . A cosmologically viable form of $f(T)$ is chosen (where $T$ is the torsion scalar in teleparallelism) in the background of homogenous and anisotropic. For our model, we take  $f(T) = T+\zeta T^{2}$ where $\zeta$ is a constant. The evolution equations are reduced to the autonomous system of differential equations by suitable transformation of variables. The behaviour of the equilibrium points is examined by calculating the eigenvalues corresponding to these equilibrium points. We get four equilibrium points for our cosmological model out of which one equilibrium point is stable, two are saddle points and one is an unstable equilibrium point.  Corresponding to equilibrium point $T_{2}$, our model is consistent with the quintessence dark energy cosmological model. Along the equilibrium points $T_{1},T_{3}$ and $T_{4}$, our model is consistent to phantom dark energy model. After that, we utilize the autonomous equations to analyse the cosmographic parameters along with the state-finder parameter. We  demonstrate  the phase plot analysis for our model.


\section{Introduction}\label{sec1}

Recently, a discrepancy has been identified between cosmological observations and Friedmann-Robertson–Walker (FRW) cosmology. Observations indicate that the universe is experiencing an era of accelerated expansion, whereas FRW cosmology, based on Einstein’s general relativity (GR), implies otherwise. This acceleration in the universe's expansion is thought to be caused by a mysterious form of energy with negative pressure, referred to as dark energy (DE). The evidence supporting the existence of this energy is derived from the observation of supernovae type Ia\cite{1,2}, anisotropies in the cosmic microwave background (CMB) measured by the Wilkinson Microwave Anisotropy Probe (WMAP)\cite{3}, and the large-scale structures of the universe\cite{4}. These observations imply that over two-thirds of our universe is made up of DE, with the remainder composed of relativistic dark matter and baryonic matter\cite{5}.\\
Modified gravity theories have attracted significant attention due to their potential to explain dark energy (DE)\cite{6}. This mysterious energy, known as DE, is dispersed nearly evenly throughout the Universe and carries negative pressure. It may be indeed identical to vacuum energy. It has been incorporated into a recent endeavor to achieve a cyclic model of the cosmos. The Universe experiences numerous eras during its maturation, each of which reflects a different value of $\omega$, the equation of state's parameter: the transition period $(\omega = -\frac{1}{3})$, the radiation-dominated era $(\omega = \frac{1}{3})$, the dust-dominated era $(\omega = 0)$, and the stiff fluid era $(\omega = 1)$. After that, it transitions towards the DE-dominated era \cite{7}. Although the cosmic constant in GR is thought to be the most straightforward choice for DE,  it has two theoretical challenges: coincidence and the cosmological constant \cite{8}.
In 1928, Einstein brought forward the doctrine of teleparallelism as an alternative way of synthesising electromagnetism and gravity into one cohesive field theory. Space-time in teleparallel is stipulated by an affine connection whose curvature is vanishing, but has torsion\cite{9}. Many scholars have been interested in the geometry of absolute parallelism (AP) in two different ways over the last two decades. The first dealt with the conceptualization of the geometry, whereas the second examined the physical application of the geometry. The primary rationale for the term "teleparallel," which refers to "parallel at a distance," is that a vector's parallel transport is dependent upon a path whose corresponding curvature is also zero. It is acknowledged that GR can really be reconstructed in a teleparallel language. The theory is known as the teleparallel equivalent of general relativity (TEGR). Numerous modified gravity theories have been postulated that would clarify the universe's acceleration, one of which is an attempt to generalise TEGR to $f(T)$ theory\cite{10}, which is comparable to that  applying GR to $f(R)$  gravity in general\cite{11}. Teleparalelism uses the Weitzenbock connection, where the torsion scalar $T$ is used instead of the curvature $R$ that is specified by the Levi-Civita connection. In modern cosmology, teleparallel Lagrangian density is expressed by the torsion scalar $T$, and has been extended to a function of $T$ called $f(T)$ gravity to take into consideration both inflation as well as the Universe's late-time accelerated expansion. \\
The Weyl tensor, local pressure anisotropy, and energy density distribution inhomogeneity are three fundamental ingredients that have been the primary objective of several investigations in scientific studies connected to the study of self-gravitating systems. The Weyl tensor or its derivative functions have been proposed as characteristics of the direction of gravitational time. This assertion is based on the notion that tidal forces cause a gravitational fluid's non-uniformity to progressively increase, implying a directed passage of time.  Notably, the connection between the density contrast and the Weyl tensor is influenced by local pressure anisotropy. A number of researchers have recently examined the Bianchi type I (BI) model where anisotropic DE is present\cite{12}. Rodrigues\cite{13} developed a BI $\Lambda$CDM cosmological model that results in an anisotropic vacuum pressure while maintaining the non-dynamical character through its DE component. The direction-dependent acceleration of the later universe could affect the earlier isotropy of the cosmos, according to Koivisto and Mota's\cite{14} alternative solution to the CMB anisotropy problem. Mota\cite{15} studied the BI cosmological model with a perfect fluid component, non-dynamical anisotropic Equation of State (EoS), and interacting DE fluid. They postulated that cosmological models with anisotropic EoS might be responsible for some of the observed anomalies in CMB and that the expansion rate of the universe would become direction dependant at late times if the EoS were anisotropic. The appropriate BI solution is computed with the assumption that the coefficient of the second derivative of $f(T)$ has certain restrictions.
 A constant torsion scalar is obtained from this solution. The field equations of $f(T)$ are applied to the anisotropic homogeneous model in this investigation. We obtain a solution whose scalar torsion is constant by applying the continuity equation. \\
 In light of these prominent aspects of the cosmos, we concentrate on the Locally Rotationally Symmetric (LRS) BI model in this work, which is a subset of the BI model. The autonomous equations are formed by this metric and the general field equations. In the subsequent steps, we introduce the anisotropic model's generalisation with torsion consideration.  The equations of motion in higher-order theories of gravity have a significant degree of non-linearity, resulting in it extremely challenging to determine the analytical as well as numerical solutions. In this article, we employ the dynamical system method to analyse the cosmic dynamics and attributes of the $f(T)$ gravity cosmological model. The corresponding stability of equilibrium points is investigated by examining the linear perturbations pertaining to the equilibrium points. Specifically, we examine their associated cosmic dynamics using actual forms of the $f(T)$ function. The dynamical system approach can be used to reinstate the cosmological model's qualitative characteristics and global dynamics. This approach is widely used for evaluating the cosmological models' general cosmic dynamics. Regardless of its inception conditions, the dynamical system technique offers a qualitative narrative of the Universe's evolution. Furthermore, this approach offers a straightforward way of obtaining the numerical solutions. In this paper, we intend to identify distinguishable evolutionary phases of the Universe possessed by the $f(T)$ gravity theory by considering a form of $f(T) = T+ \phi(T)$ . Our particular objective is to determine the stability and associated cosmographic evolution by restructuring the gravitational field equations into an autonomous system. This will allow us to contrast the $f (T)$ model's findings with those of the conventional cosmological paradigm. \\
 The contents of this article are organised as follows: in Sect. 2, we enquire about the  preliminary descriptions and equations of motion of $f(T)$ gravity. Sect. 3 is all about the Anisotropic and homogenous tetrad field equations. We study the dynamical system formulation and cosmic dynamic of $f(T)$ in Sect. 4. It has a subsections in which we analyses stability of the equilibrium points  and corresponding cosmological dynamics. In Sect. 5 we discuss  general dilemma of cosmology which has two subsection in which we discuss about the cosmography and state-finder parameter for our model using the autonomous dynamical system. And at last, in Sect. 6 we conclude.

\section{ Preliminary descriptions and equations of motion}\label{sec2}

The mathematical framework of the $f(T)$ gravity is based on the Weitzenbock geometry. The following is our nomenclature and convention. The components of the tangent space to the manifold (space-time) are illustrated by the Greek indices, while the Latin ones illustrate the components of the space-time. Generally, for space-time metrics, the line element is written as
\begin{equation} 
ds^{2} = g_{\mu\nu}dx^{\mu}dx^{\nu} = \eta_{ij}e^{i}_{\mu}e^{j}_{\nu}dx^{\mu}dx^{\nu}
\end{equation}
where $\eta_{ij}=(-1,+1,+1,+1)$ is the Minkowski metric, $e^{i}_{\mu}$ and $e^{j}_{\nu}$ are the covariant vector fields and contra-variant vector field respectively which satisfies the orthogonality and unitary conditions. The parallel vector fields in a space-time with absolute parallelism $e^{\mu}_{i}$ explain the nonsymmetric affine connection $\Gamma^{\lambda}_{\mu\nu}=e^{\lambda}_{i}e^{i}_{\mu,\nu}$ \cite{16}. The curvature tensor explained with the help of Weitzenbock connection, $\Gamma^{\lambda}_{\mu\nu}$, is identically vanishing. The contortion and torsion components are explained as $K^{\mu\nu}=-(T^{\mu\nu}\alpha-T^{\nu\mu}\alpha-T\alpha^{\mu\nu})/2$ and $T^{\alpha}_{\mu\nu} = \Gamma^{\alpha}_{\nu\mu}-\Gamma^{\alpha}_{\nu\mu} = e^{\alpha}_{a}(\partial_{\mu}e^{a}v-\partial_{\nu}e^{a}\mu)$ respectively.\\
The form of skew symmetric tensor $S^{\nu\rho}_{\mu}$ is
\begin{equation} 
S^{\nu\rho}_{\mu} = \frac{(K^{\nu\rho}_{\mu}+\delta^{\nu}_{\mu}T^{\lambda\rho}_{\lambda}-\delta^{\rho}_{\nu}T^{\lambda\nu}_{\lambda})}{2}
\end{equation}  
The torsion scalar is 
\begin{equation}
T = T^{\mu}_{\nu\rho}S^{\nu\rho}_{\mu}=\bigg(\frac{1}{4}\bigg)T^{\mu\nu\rho}T_{\mu\nu\rho}+\bigg(\frac{1}{2}\bigg)T^{\mu\nu\rho}T_{\rho\nu\mu}-T_{\mu\nu}^{\mu}T^{\rho\nu}_{\rho}.
\end{equation}   
The action of the $f(T)$ gravity is given by
\begin{equation}
S(e^{a}_{\mu},\Phi_{A}) = \int d^{4}x e\bigg[\mathcal{L}_{Matter}(\Phi_{A})+\frac{M^{2}_{PI}}{2}(f(T)-2\Lambda)\bigg]
\end{equation}
where $e = det(e^{a}_{\mu}) = \sqrt{-g}$, $\Lambda$ is the cosmological constant, $\Phi_{A}$ are the matter fields and $\mathcal{L}_{matter}$ is the Lagrangian of the matter field. In this case, $M_{PI}$ is the reduced Planck mass, which is related to the gravitational constant $G$ by $M_{PI}= \sqrt{\frac{ch}{8\pi G}}$. Likewise the $f(R)$ theory, the action of the $f(T)$ theory can be interpreted as a function of the tetrad fields $e^{a}_{\mu}$. The following equations of motion can be obtained by setting the function's variation with respect to the tetrad fields $e^{a}_{\mu}$ to be vanishing.
\begin{equation}
8\pi \mathcal{T}^{\nu}_{\mu}=S^{\nu\rho}_{\mu}T,_{\rho}f(T)_{TT}+\frac{1}{2}\delta^{\nu}_{\mu}(f(T)-2\Lambda)+[e^{-1}e^{a}_{\mu}\partial_{\rho}(e e_{a}^{\alpha}S \alpha^{\nu\rho})-T^{\alpha} \mu\lambda S \alpha^{\nu\lambda}]f(T)_{T} 
\end{equation}
where $f(T)_{T} = \frac{\partial f(T)}{\partial T}, T_{,\rho} = \frac{\partial T}{\partial x^{\rho}}, f(T)_{TT} = \frac{\partial^{2}f(T)}{\partial T^{2}}$ and $\mathcal{T}^{\nu}_{\mu}$ is the energy-momentum tensor. \\
The field equations (5) are written in terms of their partial derivatives and tetrad. It seems that these equations vary from Einstein's field equation. Following Refs \cite{17,18,19}, we can get an equation relating the scalar torsion $T$ to the Ricci scalar $R$.  These will demonstrate how teleparallel gravity and GR are equivalent.  The absence of the local Lorentz, however, means that the tetrad cannot be  omitted in favor of the metric in $Eq.(5)$. It is possible to demonstrate that the latter can be transformed into a form that roughly matches Einstein's equation.  This form serves more effectively for building $f(T)$ theory analytic solutions. To express the field equations in a covariant form, we must substitute the partial derivatives in the tensors by covariant derivatives compatible with the metric $g_{\mu\nu}$. Hence, the torsion tensor can be written as
\begin{equation}
T^{\mu\nu}\rho = e a^{\mu}(\nabla_{\nu}e^{a}\rho - \nabla_{\rho} e^{a} \nu)
\end{equation}  
After using $Eq.(6)$ the skew symmetric tensor $S^{\mu\nu}\rho$ and contorsion become 
\begin{equation}
S^{\mu\nu}\rho = \delta^{\nu}_{\rho} \eta^{ij} e^{\sigma}_{i} \nabla_{\sigma} e^{\mu}_{j} + \eta^{ij}e^{\mu}_{i}\nabla_{\rho} e_{j}^{\nu} - \delta^{\mu} _{\rho }\eta^{ij}
e_{i}^{\sigma}\nabla_{\sigma}e_{j}^{\nu}\\
\end{equation}
\begin{equation}
K^{\mu}\nu\rho = e_{a}^{\mu}\nabla_{\rho} e^{a} \nu
\end{equation}
The associated Ricci tensor become
\begin{equation}
R_{\mu\nu} = K^{\rho}_{\sigma\nu}K^{\sigma}_{\mu\rho} - K^{\rho}_{\sigma\rho}K^{\sigma}_{\mu\nu} + \nabla_{\nu}K^{\rho}_{\mu\rho} - \nabla_{\rho}K^{\rho}_{\mu\nu}
\end{equation}
After some algebraic manipulation and using above equations, we get
\begin{equation}
G_{\mu\nu} - \frac{1}{2} g_{\mu\nu}T = -S^{\sigma\rho}_{\mu}K_{\rho\sigma\nu} - \nabla^{\rho}S_{\nu\rho\mu}
\end{equation}
where $G_{\mu\nu} = R_{\mu\nu} - g_{\mu\nu}R/2$ is the Einstein tensor. After using the $Eq.(5)$, $Eq.(10)$, and the field equations of the $f(T)$ gravity , we get
\begin{equation}
f(T)_{T}G_{\mu\nu} + S_{\mu}^{\rho\nu} T_{,\rho}f(T)_{TT} + \frac{1}{2}(f(T)-Tf(T))g_{\mu\nu} = 4\pi T^{\nu}_{\mu}
\end{equation}
With a structure resembling the field equations of the $f(R)$ gravity, $Eq.(10)$ can be used as the foundation for the $f(T)$ modified gravity model. Keep in mind that the field equations are in covariant form in the  generalised case where $f(T)\neq T$. When $f(T) = T$ and constant torsion,$f(T_{0})$ , GR is recovered and the field equations are covariant and the theory is Lorentz invariant. As the principle of isotropy and homogeneity is not valid around the Big Bang, we will apply the modified field equations to an anisotropic and homogenous model and discuss them in the next sections.

\section{ Anisotropic and homogenous tetrad field equations}\label{sec3}

The spatially anisotropic and homogenous, Bianchi type I (BI), Universe which has transverse direction $x$ and two equivalent longitudinal direction $y$ and $z$ is expressed by \cite{20}
\begin{equation}
ds^{2} = dt^{2} -L_{1}(t)dx^{2} - L_{2}(t)(dy^{2}+dz^{2})
\end{equation}
where $L_{1}(t)$ and $L_{2}(t)$ are the cosmic scalar factors. For $L_{1}(t) = L_{2}(t) = a(t) $, which reduces to the flat FRW space-time. The form of BI's diagonal tetrad components is\cite{21}
\begin{equation}
[h^{i}_{\mu}] = diag(1,L_{1}(t),L_{2}(t),L_{2}(t))
\end{equation}
And the dual of the $Eq.(13)$ is of the form
\begin{equation}
[h^{\mu}_{i}] = diag(1,L_{1}(t)^{-1},L_{2}(t)^{-1},L_{2}(t)^{-1})
\end{equation}
The determinant of the matrix (13) is $h = L_{1}L_{2}^{2}$. And the components of the torsion for the tetrads are expressed as
\begin{equation}
T^{1}_{01} = \frac{\dot{L_{1}}}{L_{1}}, \hspace{0.5cm} T^{2}_{02} = \frac{\dot{L_{2}}}{L_{2}}, \hspace{0.5cm} T^{3}_{03} = \frac{\dot{L_{2}}}{L_{2}}
\end{equation} 
and the non null components of the coterminous contorsion tensor are
\begin{equation}
K^{01}_{1} = \frac{\dot{L_{1}}}{L_{1}} \hspace{0.5cm} K^{02}_{2} = \frac{\dot{L_{2}}}{L_{2}} \hspace{0.5cm} K^{03}_{3} = \frac{\dot{L_{2}}}{L_{2}}
\end{equation}
The components of the skew symmetric tensor are expressed as
\begin{equation}
S^{10}_{1} = \frac{\dot{L_{2}}}{L_{2}}, \hspace{0.5cm} S_{2}^{20} = \frac{1}{2}\bigg(\frac{\dot{L_{1}}}{L_{1}}+\frac{\dot{L_{2}}}{L_{2}}\bigg), \hspace{0.5cm} S_{3}^{30} = \frac{1}{2}\bigg(\frac{\dot{L_{1}}}{L_{1}}+\frac{\dot{L_{2}}}{L_{2}}\bigg)
\end{equation} 
 After using the $Eq.(6)-(8)$, and $Eq.(13)-(17)$, we get the torsion scalar is in the form of
 \begin{equation}
 T = -2\bigg(2\frac{\dot{L_{1}}\dot{L_{2}}}{L_{1}L_{2}} + \frac{\dot{L_{2}^{2}}}{L_{2}^{2}}\bigg)
 \end{equation}
 In the corresponding direction, Hubble's parameter is defined
 \begin{equation}
 H_{i} = \frac{\dot{a}}{a} \hspace{0.75cm} i= 1,2
 \end{equation}
 where over-head dot symbolised derivative with respect to Cosmic time. The description of the generalised Hubble parameter $H$ is expressed as follows
 \begin{equation}
 H = \frac{\dot{a}}{a} = \frac{1}{3}\bigg(\frac{\dot{L_{1}}}{L_{1}} + 2\frac{\dot{L_{2}}}{L_{2}} \bigg) = \frac{1}{3}(H_{1}+2H_{2}).
  \end{equation} 
  and here we also consider that the $L_{1} \propto L_{2}^{n}$ and $n$ is a positive constant. The value of $n$ is strictly greater than 2. The generalised Hubble parameter and the directional Hubble parameters have the following link.
  \begin{equation}
  H_{1}= n H_{2} = \bigg(\frac{3n}{n+2}\bigg)H,
  \end{equation} 
  The spatial volume is 
  \begin{equation}
  V(t) = L_{1}L_{2}^{2}
  \end{equation}
 The motion equations for the BI model emanate from
 \begin{equation}
 \frac{f}{2} + f_{T}\bigg[4\frac{\dot{L}_{1}\dot{L}_{2}}{L_{1}L_{2}} +2\bigg(\frac{\dot{L_{2}}}{L_{2}}\bigg)^{2}\bigg] = \rho\\
  \end{equation}
  \begin{equation}
 -f(T) - 4\bigg( \frac{\dot{L_{1}}\dot{L_{2}}}{L_{1}L_{2}} + \frac{\dot{L_{2}}^{2}}{L_{2}^{2}} + \frac{\ddot{L_{2}}}{L_{2}}\bigg)f_{T} - 4\frac{\dot{L_{2}}}{L_{2}}\dot{T}f_{TT} = p_{x}\\
 \end{equation}
 \begin{equation}
 -f(T) - 2\bigg( 3\frac{\dot{L_{1}}\dot{L_{2}}}{L_{1}L_{2}} + \frac{\dot{L_{2}}^{2}}{L_{2}^{2}} + \frac{\ddot{L_{2}}}{L_{2}} + \frac{\ddot{L_{1}}}{L_{1}}\bigg)f_{T} - 2\bigg(\frac{\dot{L_{1}}}{L_{1}} + \frac{\dot{L_{2}}}{L_{2}}\bigg)\dot{T}f_{TT} = p_{y} \\
 \end{equation}
 \begin{equation}
 p_{y} = p_{z}
 \end{equation}
 $Eq.(23)-(26)$ can bring down to those of TEGR when $f(T) = T$ and reduce to the FRW model when $L_{1} = L_{2} = V(t)$. 
 We consider the energy-momentum tensor $T^{\mu}_{\nu}$ as
 \begin{equation*}
 T^{\mu}_{\nu} = diag[-1,\omega_{x},\omega_{y}, \omega_{z}]\rho
 \end{equation*}
 \begin{equation}
 T^{\mu}_{\nu} = diag[-1,\omega , (\omega+ \delta), (\omega+ \delta)]\rho
 \end{equation}
 here $\rho$ is the energy density of the fluid and $p_{x},p_{y}$ and $p_{z}$ are the pressure of the fluid along $x,y$ and $z$ axes respectively. Directional Equation of state (EoS) parameter of the fluid along $x,y$ and $z$ directions are $\omega_{x}, \omega_{y}$ and $\omega_{z}$. Parameterizing the deviation from isotropy by substituting $\omega_{x} = \omega$ and incorporating the skewness parameter $\delta$ which deviates from both the $y$ and $z$ axes respectively. Skew parameter $\delta$ need not  be constant and can be a function of the Universal time $t$.\\
 For an anisotropic field, the equations of motion compatible with the BI model are derived by
 
 \begin{equation}
 \frac{f}{2} + 2f_{T} \bigg[2\frac{\dot{L}_{1}\dot{L_{2}}}{L_{1}L_{2}} + \bigg(\frac{\dot{L_{2}}}{L_{2}}\bigg)^{2}\bigg] = \rho 
 \end{equation}
 \begin{equation}
 f(T) + 4\bigg(\frac{\dot{L_{1}}\dot{L_{2}}}{L_{1}L_{2}} + \frac{\dot{L_{2}^{2}}}{L_{2}^{2}} + \frac{\ddot{L_{2}}}{L_{2}}\bigg)f_{T} + 4 \frac{\dot{L_{2}}}{L_{2}}\dot{T}f_{TT} = \omega \rho 
 \end{equation}
 \begin{equation}
 f(T) + 2 \bigg(3 \frac{\dot{L_{1}}\dot{L_{2}}}{L_{1}L_{2}} + \frac{\dot{L_{2}^{2}}}{L_{2}^{2}}+\frac{\ddot{L_{1}}}{L_{1}} + \frac{\ddot{L_{2}}}{L2} \bigg)f_{T} + 2\bigg(\frac{\dot{L_{1}}}{L_{1}} + \frac{\dot{L_{2}}}{L_{2}} \bigg)\dot{T}f_{TT} = (\omega+\delta)\rho
 \end{equation} 
here $T$ is the torsion scalar expressed as $T = -\frac{18(2n+1)}{(n+2)^{2}}H^{2}$, with $H$ representing the Hubble parameter. As we assume $f(T) = T + \phi(T)$, the $Eq(28)-(30)$ become
\begin{equation}
\frac{9(2n+1)H^{2}}{(n+2)^{2}} = \rho + \frac{2T\phi_{T}-\phi(T)}{2}
\end{equation}
\begin{equation}
\rho\delta = -\frac{(n^{2}+n-2)Tf_{T}}{(2n+1)}+ \frac{6(n-1)\dot{H}}{(n+2)}(2T\phi_{TT}+\phi_{T}+1)
\end{equation}  
where $f_{T} = \frac{df}{dT}, f_{TT} = \frac{d^{2}f}{dT^{2}}, \rho$ is the energy density and $\delta$ is the skewness parameter which is the function of time. In order to maintain our cosmological model relatively straightforward, we gravitate towards $\delta$ to be a constant. The energy density parameters are derived from $Eq(31)$ as
\begin{equation}
\Omega_{T} = \frac{(n+2)^{2}}{9(2n+1)}\bigg[\frac{2T\phi_{T}-\phi(T)}{H^{2}}\bigg] \hspace{0.95cm} \Omega_{m} = \frac{(n+2)^{2}\rho}{9(2n+1)H^{2}}
\end{equation}
We get
\begin{equation}
\Omega = \Omega_{T} + \Omega_{m} = 1
\end{equation}
Here the energy density parameter is represented by $\Omega$ which is an important concept that determines the geometry and ultimate fate of the Universe. The value of $\Omega$ lies in the range of 0 and 1 . When the value of $\Omega =1$, it exhibits that our Universe has a flat Universe and it will expand forever but at a gradually slowing rate. When the value of $\Omega < 1$, it means that our Universe has an open geometry and it will expand forever at an accelerating rate. And when the geometry of the Universe is closed and the expansion will eventually reverse into a "Big crunch", it implies that the value of $\Omega > 1$. This parameter  plays a significant role in understanding the geometry, structure, and destiny of the Universe.
  
\section{Dynamical system formulation and cosmic dynamics of $f(T) = T + \zeta T^{2}$ model}

The approach of dynamical system analysis is a tool for scrutinizing the entire dynamics of the Universe. In this approach, the qualitative behaviour of the system is studied near the equilibrium points of the model. Based on of different initial conditions, we get different solutions. The solution that is not consistent is ruled out based on the early and late time characteristics of the Universe and radiation/matter solution. A specific cosmological model may be expressed as an independent system of certain differential equations by consciously selecting the dynamical constraints which also give a prompt response to whether the model can replicate the observed expansion of the Universe. In a dynamical system to access the stability of the system. We need to transform the cosmological equations into a dynamical system by introducing some dimensionless parameters and constructing the first order autonomous system of differential equations. It is a potent way to scrutinize the Universe, where one can not get the exact solution due to the complex nature of non-linear systems of differential field equations. Let $\dot{x} = f(x)$ is the dynamical system with equilibrium point at $x_{0}$  , where $f:X \rightarrow X$ and over head dot indicate the differentiation with respect to $t \in \mathbb{R}$ and $x = (x_{1},x_{2},x_{3}...x_{n}) \in X$ be an element of $X \subseteq \mathbb{R}^{n}$. To have the equilibrium point of our dynamical system it needs to satisfy the condition $f(x) = 0$ which means that solution does not change with time. Based on  the stability, the equilibrium point can be classified as stable, unstable and saddle point. As we assume $f$ to be analytic, we can linearise the system around its equilibrium point. We have $f(x) = (f_{1}(x),f_{2}(x)...f_{n}(x))$, so that each $f_{i}(x_{1},x_{2}...x_{n})$ can be Taylor near $x_{0}$ which yields
\begin{equation}
f_{i}(x) = f_{i}(x_{0}) + \sum_{j=1}^{n}\frac{\partial f_{i}}{\partial x_{j}}(x_{0})y_{j} + \frac{1}{2!} \sum_{j,k=1}^{n}\frac{\partial^{2}f_{i}}{\partial x_{j}\partial x_{k}}(x_{0})y_{j}y_{k}+...
\end{equation}  
where $y$ is defined as $ y = x-x_{0}$. We can only consider the first partial derivatives, when we using linear stability theory. Therefore, we can defined the Jacobian matrix as follows
\begin{equation}
\mathcal{J} = \begin{bmatrix}
\frac{\partial f_{1}}{\partial x_{1}} & . & . & . & \frac{\partial f_{1}}{\partial x_{n}} \\
.\hspace{0.35cm} & .\hspace{0.35cm} & .\hspace{0.35cm}& .\hspace{0.35cm} & .\hspace{0.35cm} \\
. \hspace{0.35cm}& . \hspace{0.45cm} & .\hspace{0.35cm}& .\hspace{0.35cm} & .\hspace{0.35cm}\\
. \hspace{0.45cm}& . \hspace{0.45cm} & .\hspace{0.35cm}& .\hspace{0.35cm} & .\hspace{0.35cm}\\
\frac{\partial f_{n}}{\partial x_{1}} & . & . & . & \frac{\partial f_{n}}{\partial x_{n}}

\end{bmatrix}
\end{equation} 
Which is also called the stability matrix. We evaluate the eigenvalues of the Jacobian matrix $\mathcal{J}$  of the equilibrium point $x_{0}$ which contain  information about stability. The stability of the equilibrium point can be determined from the determinant and trace of the $\mathcal{J}$ at the equilibrium point.  If the value of determinant $|\mathcal{J}| < 0$, the eigenvalues are real and have opposite signs which implies that the equilibrium point is a saddle point. And when the value of $|\mathcal{J}| > 0$ and trace $tr (\mathcal{J})< 0$ then both the eigenvalues have negative real parts which imply that our eigenvalues are stable. When $|\mathcal{J}| >0$ and the  $tr(\mathcal{J})> 0$ then the eigenvalues are unstable. For the eigenvalues to become  nodes they need to satisfy the condition $tr(\mathcal{J})^{2} - 4 |(\mathcal{J})| > 0$ and for spirals, the eigenvalues need to satisfy the following condition $tr(\mathcal{J})^{2} - 4 |(\mathcal{J})| < 0$\cite{22}. To investigate our cosmological model in $f (T )$ gravity, we incorporate some dimensionless parameters
\begin{equation}
x = \frac{(n+2)^{2}\rho}{9(2n+1)H^{2}} \hspace{0.65cm} y = \frac{(n+2)^{2}T\phi_{T}}{9(2n+1)H^{2}} \hspace{0.65cm} z = -\frac{(n+2)^{2}\phi(T)}{18(2n+1)H^{2}}
\end{equation}
The energy balance equation is
\begin{equation}
\dot{\rho} + 3H \rho = Q
\end{equation}
here $Q$ is the interaction term related to energy transfer in the Universe. Its value can be positive or negative depending upon the nature of energy transfer. Its value is positive when the energy is transferred from dark energy to dark matter . This is required to alleviate the coincidence problem and is aligned with the second law of thermodynamics. For the time being, $Q$ is not specified, only it is considered that $Q$ does not change its sign during cosmic evolution. Because of the unknown nature of both the components (dark energy and dark matter), the precise form of the interaction can not be determined from the outset or  phenomenological requirements. For the shake of our convince, we consider the value of $Q = \alpha \rho H$, where $\alpha$ is the coupling constant and we make an assumption that its value is very small. The number of e-folding for a BI Universe is $N = \frac{1}{3}\ln(L_{1}L_{2}^{2}) = \frac{(n+2)}{3}\ln L_{2} = \bigg(\frac{n+2}{3n}\bigg)\ln L_{1}$. It is anticipated that such a system will demonstrate an accelerated expansion. In the presence of energy interaction, the evolution equations can be encapsulated as an autonomous system as follows:
\begin{equation}
x^{'} = x\bigg[\frac{(n+2)(n-1)[(1+\phi_{T})(\alpha+3)+2\phi_{TT}(\alpha-3)]-3(2n+1)x\delta}{(n+2)(n-1)(1+\phi_{T})+2T\phi_{TT}}\bigg]
\end{equation}
\begin{equation}
y^{'} = 2 T \phi_{TT} \bigg[\frac{6(n^{2}+n-2)(1+\phi_{T})-3(2n+1)x\delta}{(n+2)(n-1)(1+\phi_{T}+2T\phi_{TT})}\bigg]
\end{equation} 
\begin{equation}
z^{'} = (2z+y)\bigg[\frac{18(n^{2}+n-2)(1+\phi_{T})-9(2n+1)x\delta}{6(n+2)(n-1)(1+\phi_{T}+2T\phi_{TT})}\bigg]
\end{equation}
The decelerating parameter of the system is a dimensionless number that symbolises the rate at which the cosmos evolves concerning time. It quantifies the rate of expansion,  acceleration, or deceleration, and gives information on the general dynamics of the universe. The expression for the deceleration parameter is given as 
\begin{equation}
q = -\frac{a\ddot{a}}{\dot{a}^{2}}
\end{equation} 
\begin{equation*}
q = -1 - \frac{\dot{H}}{H^{2}}
\end{equation*}
\begin{equation}
q = -1 - \frac{18(n^{2}+n-2)(1+\phi_{T})-9(2n+1)x\delta}{6(n+2)(n-1)(1+\phi_{T}+2T\phi_{TT})}
\end{equation}
its value provides information about the dominant components of the Universe's energy content and their effects on the expansion of the cosmos. When the value of $q>0$ implies that the rate of expansion of the cosmos is slowing down. While the value of $q < 0$, it implies that our cosmos has an accelerated phase of expansion . And when the value of $q = 0$, it means that the rate of expansion is constant. The equation of state parameter $\omega_{eff}$ plays an effective role in understanding the dynamics of the cosmos expansion. Its value explains  how the component of energy density changes as the Universe  expands. When the value of $\omega_{eff}< -1$, it shows that the Universe will go through an accelerated phase of expansion of the Universe. On the other hand when $-1 < \omega_{eff} < -\frac{1}{3}$, the Universe might continue to expand indefinitely, although the rate of expansion would gradually slow down.
The formulation of $\omega_{eff}$ is as follows
\begin{equation*}
\omega_{eff} = \frac{p}{\rho}
\end{equation*}   
After using $Eq.(31)$ and $Eq.(32)$, we get
\begin{equation}
\omega_{eff} = -1 + \frac{(n+2)^{2}(n-1)[27(2n+1)+2(25n+1)-9(4n+5)y]+18[(2n+1)x \delta-2(n^{2}+n-2)(1+\phi_{T})}{9(n+2)^{2}(n-1)(2n+1)(z-1)}
\end{equation}
As we assume the specific form of $f(T) = T+ \zeta T^{2} $, which is quadratic and dependent on $T$.  The value of $\phi_{T},\phi_{TT}$ and ($2T\phi_{TT} + \phi_{T}+1)$ are $2\zeta T , 2\zeta$ and $(1-\frac{3}{2}y)$ respectively. We select this model because quadratic model preserves the second order nature of the field equation. Geometrically, it enhances the role of torsion in strong gravitational fields while maintaining mathematical tractability. Cosmologically, this form can naturally generate accelerated expansion without a cosmological constant and provides a smooth transition between cosmic epochs, with the $T^{2}$ term dominating in the early universe and becoming subdominant later \cite{23}. The autonomous dynamical system for our model in $f(T)$ gravity become
\begin{equation}
x^{'} = x\bigg[\frac{(n+2)(n-1)(2-y)(\alpha+3)+8\zeta(\alpha-3)-6(2n+1)x\delta}{(n+2)(n-1)(2-3y)}\bigg]
\end{equation}
\begin{equation}
y^{'} = 6y\bigg[\frac{(n^{2}+n-2)(y+2)+(2n+1)x\delta}{(n+2)(n-1)(2-3y)}\bigg]
\end{equation}
\begin{equation}
z^{'} = 3(2z+y)\bigg[\frac{(2n+1)x\delta+(n^{2}+n-2)(y-2)}{(n+2)(n-1)(3y-2)}\bigg]
\end{equation}
And the value of decelerating parameter $(q)$ and the Equation of state parameter $(\omega_{eff})$ become
\begin{equation}
q = -1 + \frac{3(n^{2}+n-2)(y-2)+3(2n+1)x\delta}{(n+2)(n-1)(3y-2)}
\end{equation}
And
\begin{equation}
\omega_{eff} = -1 + \frac{(n+2)^{2}(n-1)[27(2n+1)+2(25n+11)-9(4n+5)y]+18[(2n+1)x\delta+(n^{2}+n-2)(y-2)]}{9(n+2)^{2}(n-1)(2n+1)(z-1)}
\end{equation}
As our dynamical system is a non-linear dynamical system and to analyses the stability of the model, we have to calculate the equilibrium points for our cosmological model by substituting the values of $x^{'}=0, y^{'}=0$ and $z^{'}=0$. And to analyse the stability of the model by calculating the eigenvalues of the stability matrix, which is assessed at the equilibrium points $(x_{0},y_{0},z_{0})$.
\begin{equation}
\mathcal{J} = \begin{bmatrix}
\frac{\partial f}{\partial x} \hspace{0.35cm}& \frac{\partial f}{\partial y} \hspace{0.35cm}& \frac{\partial f}{\partial z}\\
\frac{\partial g}{\partial x} \hspace{0.35cm}& \frac{\partial g}{\partial y} \hspace{0.35cm}& \frac{\partial g}{\partial z} \\
\frac{\partial h}{\partial x} \hspace{0.35cm}& \frac{\partial h}{\partial y} \hspace{0.35cm}& \frac{\partial h}{\partial z} 
\end{bmatrix}
\end{equation}
where $f(x,y,z) = x^{'} , g(x,y,z) = y^{'}$ and $h(x,y,z) = z^{'}$\\
Our cosmological model, we obtain four equilibrium points i.e.\\
$\bullet$ Equilibrium point $T_{1}$ (0,0,0)\\
$\bullet$ Equilibrium point $T_{2}$ $(0,2,\gamma)$\\
$\bullet$ Equilibrium point $T_{3}$ $\bigg(-\frac{8\zeta}{(2n+1)\delta}, \frac{2(n^{2}+n-2+4\zeta)}{n^{2}+n-2},\gamma\bigg)$\\
$\bullet$ Equilibrium point $T_{4}$ $\bigg(\frac{(n^{2}+n-2)(3+\alpha)-4\zeta(3-\alpha)}{3\delta (2n+1)},0,0 \bigg)$\\

\subsection{The stability of equilibrium points  and corresponding cosmos dynamics}
These equilibrium points and the cosmological quantities corresponding to these equilibrium points are securities in Table-I. The behaviour of these equilibrium points are analysed by calculating the eigenvalues of stability matrix, which we show in Table-II. From Table-I, we come to the conclusion that the equilibrium points $T_{1}$ and $T_{2}$ always exist, while the equilibrium points $T_{3}$ and $T_{4}$ exist when the value of $\delta \neq 0$.\\
\begin{enumerate}
  \item $T_{1}$ : This equilibrium point  always exist. And the eigenvalues corresponding to this equilibrium point which we obtained from the stability matrix are $(-6,6,(3+\alpha)-\frac{4\zeta(3-\alpha)}{(n^{2}+n-2)})$. As we  observe that the eigenvalues corresponding to this equilibrium point are of opposite sign. Hence we conclude that the equilibrium point $T_{1}$ is saddle point as shown in Fig. 1 . The value of decelerating parameter at this equilibrium point is $q = 2$ which resemble that our cosmos is decelerated. At this equilibrium point, the value of $\omega_{eff} =-\frac{346}{45} < -1$ which indicate that our model is phantom dark energy model. \\
  \item $T_{2}$: This point always exist have the eigenvalues $0,c_{1},c_{2}$ which we calculated from the stability matrix where $c_{1} = \frac{-6(n^{2}+n-2)-4\zeta(\alpha-3)-\sqrt{[6(n^{2}+n-2)-4\zeta(\alpha-3)]^{2}-4(24(n^{2}+n-2)(\alpha+3)\zeta+6\delta \zeta(2n+1)(\alpha-3))}}{4(n-1)(n+2)}$ ,and  $c_{2} = \frac{-6(n^{2}+n-2)-4\zeta(\alpha-3)+\sqrt{[6(n^{2}+n-2)-4\zeta(\alpha-3)]^{2}-4(24(n^{2}+n-2)(\alpha+3)\zeta+6\delta \zeta(2n+1)(\alpha-3))}}{4(n-1)(n+2)}$. All the eigenvalues are of the negative sign which implies that our equilibrium point is the stable equilibrium point as shown in Fig. 1 . The decelerating parameter $q = -1$ which indicate that our cosmos is in the epoch of the accelerated expansion. And $\omega_{eff} = -\frac{14}{25} $ which is greater that -1 and less that $-\frac{1}{3}$ which indicate that our model is in quintessence dark energy model. 
  \item $T_{3}$: The point is exist when the value of $\delta \neq 0$. The eigenvalues corresponding to this equilibrium point are $(\frac{24\zeta}{(n+2)(n-1)+6\zeta},d_{1},d_{2})$, where $d_{1}=\frac{-6\delta(n^{3}+30n^{2}-4n-8)-\sqrt{16\delta^{2}(242n^{4}-48n^{3}-87n^{2}+12n+12)+3840\zeta \delta^{2}(12n^{2}-3n-2)}}{4\delta(n+2)(n-1)(2n+1)(n^{2}+n-2+6\zeta)^{2}}$ ,and $d_{2}=\frac{-6\delta(n^{3}+30n^{2}-4n-8)+\sqrt{16\delta^{2}(242n^{4}-48n^{3}-87n^{2}+12n+12)+3840\zeta \delta^{2}(12n^{2}-3n-2)}}{4\delta(n+2)(n-1)(2n+1)(n^{2}+n-2+6\zeta)^{2}}$. As we can observe that some eigenvalues are of positive sign while some are of negative sign which implies that our point is the saddle point as shown in Fig. 1  which is attractive in one direction and repulsive in others. The value of decelerating parameter $q$ and Equation of State parameter $\omega_{eff}$ are $-\frac{3}{2}$ and $-1.374$ respectively. As the value of $\omega_{eff}<-1$ which implies that our model is phantom dark energy model. The value of $q = -3/2$ which means that our Universe is accelerated one. 
  \item $T_{4}$: This equilibrium point exist when the value of $\delta \neq 0$. $(\frac{(\alpha-3)(n^{2}+n-2+4\zeta)}{(n+2)(n-1)},(\alpha+3)+\frac{4\zeta(\alpha-3)}{(n+2)(n-1)}, (\alpha+9)+\frac{4\zeta(\alpha-3)}{(n+2)(n-1)}) $ are the eigenvalues corresponding to this point. As we observe that all the eigenvalues are of positive sign which conclude that our point is unstable point which is shown in the Fig.1 . Corresponding to this point the decelerating parameter become $q = 2$ which means that our model is decelerated one. And the value of $\omega_{eff} = -6.680$ which is less than -1 which again implies that our model is phantom dark energy model. \\
      Hence for our model, we get four equilibrium points out of which one equilibrium point is stable , two are saddle points and one is unstable equilibrium point. Equilibrium points $T_{1},T_{3}$ and $T_{4}$  correspond to the phantom dark energy model while the equilibrium point $T_{2}$ correspond to the quintessence dark energy model. \\ 
\end{enumerate}
\begin{table}
\centering
\caption{ The cosmological quantities at the equilibrium point of $f(T) =T+\zeta T^{2} $ model}
\begin{tabular}{|p{1cm}|p{1cm}|p{1cm}|p{2.5cm}|p{3cm}|p{3cm}|p{2cm}|}
\hline
Points & $q$ & $\omega_{eff}$ & $\Omega_{T}$ & $\Omega_{m}$ & $r$ & $s$ \\
\hline
$T_{1}$ & 2 & $-\frac{346}{45}$ & 0 & 1& $\frac{9}{2}$ & $\frac{7}{9}$ \\
\hline
$T_{2}$ & -1 & $-\frac{14}{25}$ & 1&0 & 0 & $\frac{1}{4}$ \\
\hline
$T_{3}$ & $-\frac{3}{2}$ & -1.374 & 2.33  & -1.33 & 3 & $-\frac{1}{3}$ \\
\hline
$T_{4}$ & 2 & -6.680 & 0 & 1 & 2.24 & $\frac{11}{40}$ \\
\hline

\end{tabular}\\
\end{table}

\begin{figure}[h!]
\includegraphics[height=3in]{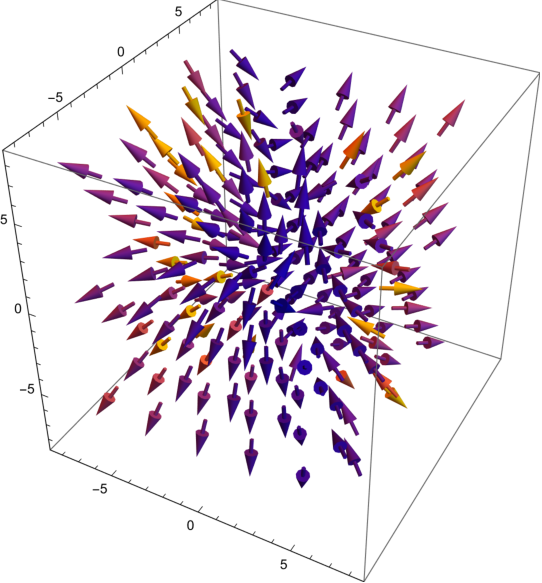}

{{Figure 1:} 3D Phase space analysis of $f(T)$ gravity for our cosmological model .}
\end{figure}
\begin{table}[h!]
\caption{ Eigenvalues for the model}
\centering

\begin{tabular}{|p{1cm}|p{3cm}|p{3cm}|p{3cm}|p{2cm}|}
\hline
Points & $m_{1}$ & $m_{2}$ & $m_{3}$ &  Nature \\
\hline
$T_{1}$ & -6 & 6 & $(3+\alpha)\frac{-4\zeta (3-\alpha)}{n^{2}+n-2}$   & Saddle \\
\hline
$T_{2}$ & 0 & $c_{1}$ & $c_{2}$& Stable \\
\hline
$T_{3}$ & $\frac{24\zeta}{n^{2}+n-2+6\zeta}$ & $d_{1}$ & $d_{2}$   & Saddle \\
\hline
$T_{4}$ & $\frac{(\alpha-3)(n^{2}+n-2+4\zeta)}{(n+2)(n-1)}$ & $(\alpha+3)+\frac{4\zeta(\alpha-3)}{(n+2)(n-1)}$ & $(\alpha+9)+\frac{4\zeta(\alpha-3)}{(n+2)(n-1)}$ & Unstable\\
\hline
\end{tabular}
\caption*{where $c_{1} = \frac{-6(n^{2}+n-2)-4\zeta(\alpha-3)-\sqrt{[6(n^{2}+n-2)-4\zeta(\alpha-3)]^{2}-4(24(n^{2}+n-2)(\alpha+3)\zeta+6\delta \zeta(2n+1)(\alpha-3))}}{4(n-1)(n+2)} ,\\  c_{2} = \frac{-6(n^{2}+n-2)-4\zeta(\alpha-3)+\sqrt{[6(n^{2}+n-2)-4\zeta(\alpha-3)]^{2}-4(24(n^{2}+n-2)(\alpha+3)\zeta+6\delta \zeta(2n+1)(\alpha-3))}}{4(n-1)(n+2)} ,\\ d_{1}=\frac{-6\delta(n^{3}+30n^{2}-4n-8)-\sqrt{16\delta^{2}(242n^{4}-48n^{3}-87n^{2}+12n+12)+3840\zeta \delta^{2}(12n^{2}-3n-2)}}{4\delta(n+2)(n-1)(2n+1)(n^{2}+n-2+6\zeta)^{2}}, \\ d_{2}=\frac{-6\delta(n^{3}+30n^{2}-4n-8)+\sqrt{16\delta^{2}(242n^{4}-48n^{3}-87n^{2}+12n+12)+3840\zeta \delta^{2}(12n^{2}-3n-2)}}{4\delta(n+2)(n-1)(2n+1)(n^{2}+n-2+6\zeta)^{2}}$.}
\end{table}

 \section{General dilemma}
 \subsection{Cosmography}
 Cosmographic quantities like the Hubble parameter $(H)$, deceleration parameter $(q)$, and jerk  parameter $j$ can be used to study the cosmic evolution of a model. These cosmological parameters are defined as \cite{24,25}.
 \begin{equation}
 j = \frac{\dddot{a}}{aH^{3}}   \hspace{1.5cm} q = - \frac{\ddot{a}}{aH^{2}}
 \end{equation}
where overhead dot represent the derivative with respect to the cosmic time.  The value of decelerating parameter $q$ can be greater than zero or less than zero. When the $q>0$ which means the accelerating expansion of the cosmos. When the value of $q<0$ then we have decelerating expansion of the cosmos. When the value of $q = -1$. it correspond to de-sitter expansion. In the existing dynamical system, the time derivative is associated to  $N = \frac{1}{3}\ln(L_{1}L_{2}^{2}) = \frac{(n+2)}{3}\ln L_{2} = \bigg(\frac{n+2}{3n}\bigg)\ln L_{1}$. and by using $Eq.(48)$ we can calculate the decelerating parameter for our model. The jerk parameter can be rewrite as \\
\begin{equation}
j = q(1+2q)-\frac{dq}{dN}
\end{equation}
\begin{equation}
\begin{split}
j = \frac{3(2n+1)(2-y)(\alpha-3)\delta x}{(n+2)(n-1)(2-3y)^{3}}-\frac{3(2n+1)x\delta}{(n+2)(n-1)(3y-2)}-\frac{2(2n+1)xy\delta(38-15y)}{(n+2)(n-1)(2-3y)^{3}} \\
+ \frac{3(y-2)(3y^{2}+y+6)}{(3y-2)^{3}}+\frac{3(2n+1)\delta x}{(n+2)^{2}(n-1)^{2}(3y-2)^{3}}[3x\delta(2n+1)(3y-2)-4]
\end{split}
\end{equation}
The value of jerk parameter and decelerating parameter corresponding to these equilibrium points are shown in Table-I. 
\subsection{State-finder parameter examination}
The state-finder parameter ${(r,s)}$ is a geometrical examination which enables us to distinguish the properties of dark energy in a model in an autonomous way \cite{26,27}. These diagnostic characteristics belong into one of two categories: Chaplygin gas and quintessence. When the state-finder parameter lies in the range $(-1,-\frac{1}{3})$, it indicates the quintessence kind of dark energy. The unification of  decelerating phase and accelerating phase is done by Chaplygin gas of the cosmos in the model. When the value of $r=1,s=0$, it corresponds to the $\Lambda$CDM cosmological model. The mathematical expansion of the jerk parameter $j$ is same as that of state-finder $r$. The state-finder $s$ is defined as
\begin{equation}
s= \frac{r-1}{3(q-\frac{1}{2})}
\end{equation}
By using $Eq.(48)$ and $Eq.(53)$, we calculate the parameters at the equilibrium points. And the summary of them are shown in Table-I.
\section{Conclusion} 
 This study examines the locally rotationally symmetric Bianchi type-I model within the context of the $f(T)$ theory of gravity, where $T$ stands for the torsion scalar. We focus on the examination of the LRS Bianchi type-I Universe in an innovative modified $f(T)$ gravity. In this work, we consider a cosmological model in $f(T)$ gravity. We directed a form of the $f(T)$ as $f(T)=T+\zeta T^{2}$. For our model we perform the dynamical system analysis by converting the evolution equations in to the autonomous system of differential equations. This is usually done by substituting the dimensionless parameter in to the equations. The variables which are used for defining the dynamical system comprise of the phase space and the equilibrium points. For our model, we get four equilibrium points which are $T_{1}$ (0,0,0), $T_{2}$ $(0,2,\gamma)$, $T_{3}$ $\bigg(-\frac{8\zeta}{(2n+1)\delta}, \frac{2(n^{2}+n-2+4\zeta)}{n^{2}+n-2},\gamma\bigg)$ and $T_{4}$ $\bigg(\frac{(n^{2}+n-2)(3+\alpha)-4\zeta(3-\alpha)}{3\delta (2n+1)},0,0 \bigg)$. Out of these four equilibrium point only one equilibrium point$(T_{2})$ is stable, two$(T_{1},T_{3})$ are saddle points and  $T_{4}$ is unstable equilibrium point. The equilibrium point $T_{2}$ corresponds to quintessence dark energy cosmological model while the equilibrium points $T_{1},T_{3}$ and $T_{4}$ correspond to the phantom dark energy cosmological model. We use the autonomous dynamical system to analyses the cosmography and state-finder parameter. And their analysis is shown in Table-I.  We come to the conclusion that our  model LRS Bianchi type-I of $f(T)$ gravity in presence of interaction of energy may effectively explain the acceleration of the Universe. All the cosmological parameters, cosmography and state-finder parameter are derived from the model address. Although the fact that we have theoretically produced and investigated the LRS Bianchi type-I Universe within a modified gravity framework, our results are consistent with the results of the observation. Future studies comparing these models to empirical data from BAO, Pantheon, and Hubble would be interesting. We would like to provide an update on a few of these upcoming investigations.\\
 
 $\textbf{Data Availability}$ No data is associated in the manuscript. \\
 $\textbf{Funding sources}$ This research did not receive any specific grant from funding agencies in the public, commercial, or not-for-profit sectors.\\


\begin{thebibliography} {99}
\bibitem{1} Cervantes-Cota J L and Smoot G 2011 AIP Conf. Proc. $\textbf{1396}$ 28.
\bibitem{2}  Rebolo R et al. 2004 Mon. Not. Roy. Astron. Soc. $\textbf{353}$ 747.
\bibitem{3} Bennett C L 2003 Astrophys. J. Suppl. $\textbf{148}$ 1.
\bibitem{4} Hawking E 2003 Mon. Not. Roy. Astr. Soc. $\textbf{346}$ 78.
\bibitem{5} Hinshaw G 2009 Astrophys. J. Suppl. $\textbf{180}$ 225.
\bibitem{6} Rathore, S., Singh, S.S., Muhammad, S. et al. Phase space properties of cosmological models in $f(Q, T)$ gravity. Eur. Phys. J. C $\textbf{84}$, 1108 (2024).
\bibitem{7} Rathore, S., Singh, S.S. Stability aspects of an LRS Bianchi type-I cosmological model in $f(Q)$ gravity. Gen Relativ Gravit $\textbf{56}$, 25 (2024).
\bibitem{8} Viennot D J and Vigoureux M 2009 Int. J. Theor. Phys. $\textbf{48}$ 2246
\bibitem{9} Amit Samaddar and Surendra Sanasam 2024 Phys. Scr. $\textbf{99}$ 035219
\bibitem{10} Cai, Yi-Fu et al. “$f(T)$ teleparallel gravity and cosmology.” Reports on Progress in Physics $\textbf{79}$ (2015).
\bibitem{11} De Felice, A., Tsujikawa, S. f(R) Theories. Living Rev. Relativ. $\textbf{13}$, 3 (2010).
\bibitem{12} Rathore, S., Singh Dynamical system analysis of interacting dark energy in LRS Bianchi type I cosmology S.S.: Sci. Rep. $\textbf{13}$, 13980 (2023). 
\bibitem{13} Rodrigues D C 2008 Phys. Rev. D $\textbf{77}$ 023534.

\bibitem{14} Koivisto T and Mota D F 2008 Astrophysical J. $\textbf{679}$ 1.
\bibitem{15}  Mota D et al. 2007 Mon. Not. Roy. Astron. Soc. $\textbf{382}$ 793.
\bibitem{16} Weitzenbock R 1923 Invariance Theorie (Nordhoff: Gronin-gen).
\bibitem{17} Li B, Sotiriou T P and Barrow J D 2011 Phys. Rev.D $\textbf{83}$ 064035.
\bibitem{18} Sotiriou T P, Li B and Barrow J D 2011 Phys. Rev. D $\textbf{83}$ 104030.
\bibitem{19} Li B, Sotiriou T P and Barrow J D 2011 Phys. Rev. D $\textbf{83}$ 104017.
\bibitem{20} Sharif M and Zubair M 2010 Astrophys. Space Sci. $\textbf{330}$ 399.
\bibitem{21} Sharif M and Rani S 2011 Mod. Phys. Lett. A $\textbf{26}$ 1657.
\bibitem{22} Strogatz,S.H.:Nonlinear Dynamics and Chaos: With Applications to Physics, Biology, Chemistry and Engineering. Westview Press, Boulder(2001).
\bibitem{23} Ran Chen, Yi-Ying Wang, Lei Zu, Yi-Zhong Fan Prospects of constraining $f(T)$ gravity with the third-generation gravitational-wave detectors Phys. Rev. D $\textbf{109}$, 024041 (2024).
\bibitem{24} A.R. Lalke, G.P. Singh, A. Singh, Eur. Phys. J. Plus $\textbf{139}$, 288 (2024).
\bibitem{25}  S. Mandal, A. Singh, R. Chaubey, Int. J. Geom. Methods Mod. Phys. $\textbf{20}$, 2350084 (2023).
\bibitem{26} U. Alam, V. Sahni, T.D. Saini, A.A. Starobinsky, Mon. Not. R. Astron. Soc. $\textbf{344}$, 1057 (2003).
\bibitem{27} R. solanki , P.K. Sahoo Statefinder Analysis of Symmetric Teleparallel Cosmology Annalen der Physik, $\textbf{534(6)}$ (2022).







\end{thebibliography}
\end{document}